\documentclass[12pt]{article}
\usepackage{amsmath,amssymb}
\usepackage{graphicx}
\newcommand{\eqdef}{\stackrel{\text{def}}{=}}

\newcommand{\n}{\nonumber\\}

\newcommand{\ignore}[1]{}
\numberwithin{equation}{section}
\newcommand{\Romannumeral}[1]{\uppercase\expandafter{\romannumeral#1}}

\newtheorem{theo}{\bf Theorem}[section]

\newtheorem{rema}[theo]{\bf Remark}

\allowdisplaybreaks[4]

\begin{document}

\baselineskip=20pt

\newcommand{\preprint}{
\vspace*{-20mm}
   \begin{flushright}\normalsize   
  \end{flushright}}
\newcommand{\Title}[1]{{\baselineskip=26pt
  \begin{center} \Large \bf #1 \\ \ \\ \end{center}}}
\newcommand{\Author}{\begin{center}
  \large \bf  Ryu Sasaki\end{center}}
\newcommand{\Address}{\begin{center}
      Department of Physics, Tokyo University of Science,
     Noda 278-8510, Japan
   \end{center}}
\newcommand{\Published}[1]{\begin{center}
 {\small \sf Published} \vspace{1mm}\\
   {\large \sf #1}
  \end{center}}

\preprint
\thispagestyle{empty}

\Title{Exactly solvable piecewise analytic double well potential $V_{\text D}(x)=\text{min}[(x+d)^2,(x-d)^2]$      
     and its dual \\
     single well  potential $V_{\text S}(x)=\text{max}[(x+d)^2,(x-d)^2]$ }

\Author

\Address
\vspace{1cm}

\begin{abstract}
By putting two harmonic oscillator potential $x^2$ side by side with a separation $2d$, 
two exactly solvable piecewise analytic quantum systems with a free parameter $d>0$ are obtained.
Due to the mirror symmetry, their eigenvalues $\{E\}$ for the even and odd parity sectors are determined exactly as
the zeros of certain combinations of the confluent hypergeometric function ${}_1F_1$ of $d$ and $E$, which are common
to $V_{\text D}$ and $V_{\text S}$ but in  two different branches. The eigenfunctions are the piecewise square integrable 
combinations of ${}_1F_1$, the so called $U$ functions. By comparing the eigenvalues and eigenfunctions
for various values of the separation $d$, vivid pictures unfold showing  the tunneling effects between the two wells.
\end{abstract}

\Published{Journal of Mathematical Physics {\bf 64} (2023) 022102}
\section{Introduction}
\label{sec:intro}
Double well potentials in quantum mechanics are discussed in various contexts, {\em e.g.}
tunneling or spontaneous symmetry breaking, etc.
Most commonly studied are the quartic potentials $V_{\text Q}(x)=x^4-ax^2+bx $, ($a>0$),
whose eigenfunctions are quite complicated and the system is far from exactly solvable.
Some solvable examples are the square double well and the double Dirac delta potential.
Recently Miloslav Znojil  introduced an interesting double  and single well potentials 
\eqref{dw1}, \eqref{sw1} \cite{znojil22}.
They are just  mirror symmetrically separated harmonic oscillator potentials, which are not analytic at the origin
but exactly solvable due to the harmonic oscillator nature. One motivation of this paper is to supplement the 
pioneering work of Znojil. 

Another profound motivation is to disseminate the possibility to enlarge the list 
of exactly solvable quantum mechanical systems by combining the technique of constructing mirror symmetric 
non-analytic solvable potentials \cite{znojil22, sz1, sz2, sz3} with other known methods of solvability
such as the  factorisation \cite{infhul,susyqm}, shape-invariance \cite{genden}, 
the exceptional and multi-indexed polynomials \cite{gomez, quesne,os16,os25}, 
non-polynoial extensions \cite{os31}, etc,  in particular, the Krein-Adler deformations \cite{krein, adler,os29}.
By incorporating the present solvable model construction method to the simplest Krein-Adler deformation
of the harmonic oscillator potential \cite{dubov}, a new double well potential is proposed in \S\ref{sec:comm}
\eqref{DSdubov}. It is expected to be the breakthrough point for constructing 
a multitude of similarly exactly solvable potentials.

\begin{figure}[h]
    \begin{tabular}{cccc}
      \begin{minipage}[b]{0.45\hsize}
        \centering
        \includegraphics[keepaspectratio, scale=0.8]{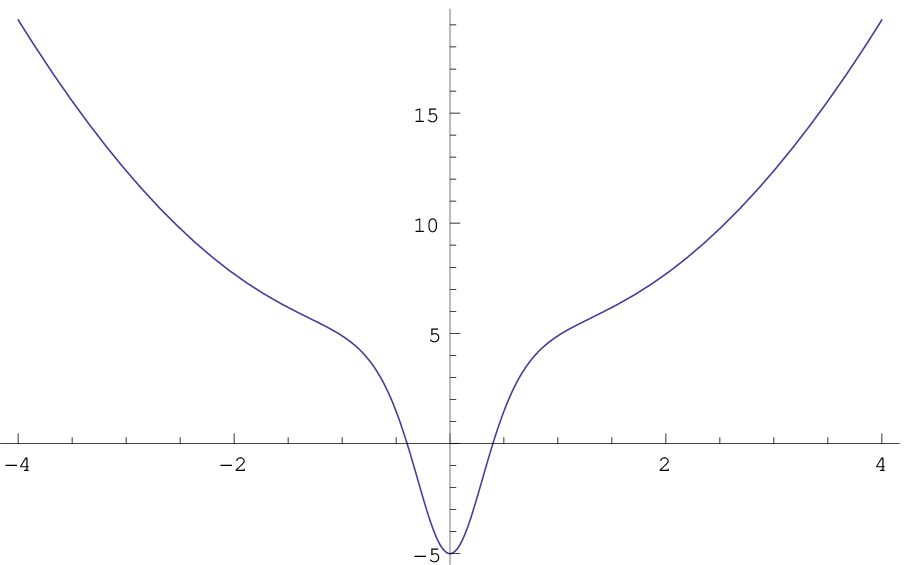}
        \caption{$V_{\rm KA}(x)$}
        \label{fig:VKA}
      \end{minipage} &
      \begin{minipage}[b]{0.45\hsize}
        \centering
        \includegraphics[keepaspectratio, scale=0.8]{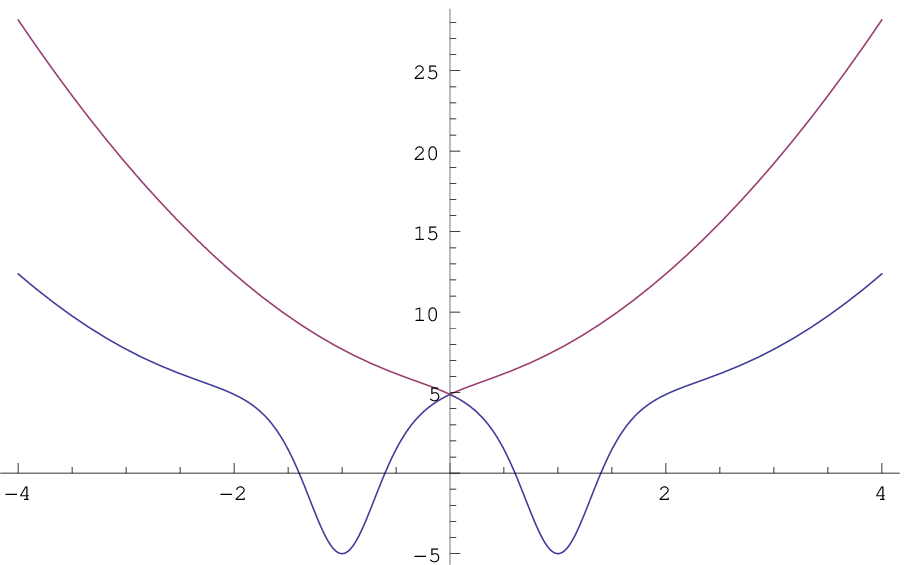}
        \caption{$V_{\rm KAD}(x)$ blue, $V_{\rm KADS}(x)$ red, $d=1$}
        \label{fig:VKADS}
      \end{minipage}
         \end{tabular}
 \end{figure}

 The present paper is prepared in a plain style so that non-experts can easily understand.
 This paper is organised as follows.
In section two, after a brief introduction of the double and single well potentials, the simplification of the
connection conditions in mirror symmetric potentials due to the separation into the even and odd sectors is 
recapitulated. 
The symmetric relationships between the connection conditions of the double and single well potentials
are stated as {\bf Theorem \ref{theo:2-1}}. A simple lower bound of the eigenvalues of the single well potential
is mentioned as {\bf Remark \ref{rem:bound}}.
In section three, elementary polynomial type solutions are briefly surveyed. 
The connection conditions determining the eigenvalues of the even and odd sectors are expressed as the zeros 
of the Hermite polynomials and the derivatives in {\bf Theorem \ref{theo:3-1}}.
The polynomial type eigenfunctions are displayed in {\bf Theorem \ref{theo:3-2}} 
together with the explicit expressions of the connection conditions in Tables \ref{tab:evenpara} and  \ref{tab:oddpara}.
The relationship with the results in \cite{znojil22} is remarked. 
The upper bound of the greatest zeros of the Hermite polynomials is mentioned 
in connection with {\bf Remark \ref{rem:bound}}.
In \S\ref{sec:nonpoly}, starting with  the Kummer differential equation, 
the piecewise square integrable combination of the confluent hypergeometric functions are introduced.
{\bf Theorem \ref{theo:4-1}} states that the eigenvalues are obtained as the zeros of the connection conditions
for the even and odd sectors.
The `duality' of the connection conditions for the double and single well potentials is alluded 
in {\bf Remark \ref{rem:otherbr}}.
Seven lowest eigenvalues for the double and single well potentials for  a few small values of $d$'s are 
shown in Tables \ref{tab:Dweig} and \ref{tab:Sweig}. Several graphs of some lower eigenfunctions are shown 
in Figures \ref{fig:1}--\ref{fig:4}. A simple interpretation of the tunneling effects on the even and odd 
sector eigenvalues are presented.
In \S\ref{sec:poly2} it is shown that various quantities and expressions in \S\ref{sec:kummer} are simplified for
odd integer eigenvalues.
In  section five, after a very brief  summary, two  new piecewise analytic double and single well potentials are
proposed \eqref{linpot}, \eqref{adlerH2fun} together with the graphs of the potentials 
in Figure \ref{fig:VKA} and \ref{fig:VKADS}.

\section{Mirror symmetric and piecewise analytic potential}
\label{sec:mirror}

Here we explore a new exactly solvable double well potential $V_{\text D}(x)$ 
and its dual single well potential $V_{\text S}(x)$  with $d>0$, 
\begin{align} 
 V_{\text D}(x)=\text{min}[(x+d)^2,(x-d)^2]=
\left\{
\begin{array}{cc}
(x-d)^2  &   x\ge0   \\[2pt]
(x+d)^2  &   x\le0    
\end{array}
\right.,\qquad V_{\text D}(x)=V_{\text D}(-x),
\label{dw1}\\[4pt]
V_{\text S}(x)=\text{max}[(x+d)^2,(x-d)^2]=
\left\{
\begin{array}{cc}
(x+d)^2  &   x\ge0   \\[2pt]
(x-d)^2  &   x\le0    
\end{array}
\right., \qquad V_{\text S}(x)=V_{\text S}(-x),
\label{sw1}
\end{align}
which were recently
introduced by Miloslav Znojil in a pioneering work \cite{znojil22}.
This paper will be cited as I hereafter.
For $d=0$, $V_{\text D}$ and $V_{\text S}$ reduce to  the well known  harmonic oscillator 
$V_{\text D}(x)=V_{\text S}(x)=x^2$.
Since the min, max definitions of $V_{\text D}(x)$ and  $V_{\text S}(x)$ are obviously symmetric 
with $d\leftrightarrow -d$, we have restricted to $d>0$ 
and the $d$-dependence of the potentials, the wavefunctions, eigenvalues 
etc is usually suppressed for the simplicity of presentation.

These potentials are obviously mirror symmetric $V(x)=V(-x)$ and analytic on either half line  $x>0$ and $x<0$
and the non-analyticity occurs only at the origin $x=0$.
That is, the wavefunctions on either half line  are analytic functions.
That is, their wavefunctions $\psi_{\text D}(x)$ and  $\psi_{\text S}(x)$ of the Schr\"odinger equations
\begin{align}
 -\frac{d^2\psi_{\text D}(x,E)}{dx^2}+V_{\text D}(x)\psi_{\text D}(x,E)=E\psi_{\text D}(x,E),\quad 
  -\frac{d^2\psi_{\text S}(x,E)}{dx^2}+V_{\text S}(x)\psi_{\text S}(x,E)=E\psi_{\text S}(x,E),
 \label{Seq1}
\end{align} 
 are piecewise analytic
 \begin{align*}
\psi_{\text D}(x,E)=
\left\{
\begin{array}{cc}
\psi_{\text D}^{(+)}(x,E) &   x>0   \\[2pt]
\psi_{\text D}^{(-)}(x,E)  &   x<0    
\end{array}
\right., \qquad 
\psi_{\text S}(x,E)=
\left\{
\begin{array}{cc}
\psi_{\text S}^{(+)}(x,E) &   x>0   \\[2pt]
\psi_{\text S}^{(-)}(x,E)  &   x<0    
\end{array}
\right..
 \end{align*}
 Let us assume that the above wavefunctions are piecewise square integrable for both D and S,
 \begin{equation*}
\int_0^\infty\left(\psi^{(+)}(x,E)\right)^2dx<\infty,\quad 
\int_{-\infty}^0\left(\psi^{(-)}(x,E)\right)^2dx<\infty.
\end{equation*}
 This selects one solution in the two-dimensional solution space of the above Schr\"odinger equations
 \eqref{Seq1} for generic $E$.
 Like other one-dimensional quantum mechanical systems with piecewise analytic potentials, 
we require  the continuity of the wavefunctions and their first derivatives
\begin{equation}
\psi^{(+)}(0,E)=\psi^{(-)}(0,E),\quad \frac{d\psi^{(+)}}{dx}(0,E)=\frac{d\psi^{(-)}}{dx}(0,E).
\label{gencond}
\end{equation}
{\em These select  the eigenvalues $\{E_n\}$, $n=0,1,\ldots$, since the continuous 
wavefunctions are square integrable eigenfunctions $\{\psi(x,E_n)\}$},
\begin{equation*}
\int_{-\infty}^\infty\psi(x,E_n)^2dx<\infty,\qquad n=0,1,\ldots.
\end{equation*}

Thanks to the mirror symmetry, this solution process is simplified extensively as demonstrated 
in other similar examples, $V(x)=-g^2{\rm exp}(-|x|)$ \cite{sz1}, $V(x)=g^2{\rm exp}(2|x|)$ \cite{sz2}  and
symmetric Morse potential \cite{sz3}, etc.
Due to the mirror symmetry of the potentials 
 the wavefunctions are split into the even and odd parity sectors
 \begin{align}
{\rm even:}& \quad \psi^{(+)}(x,E)=\psi^{(-)}(-x,E),  \qquad \  \
 {\rm odd:}& \quad \psi^{(+)}(x,E)=-\psi^{(-)}(-x,E).
 \label{evoddcond}
 \end{align}
  
The continuity of the wavefunctions and their first derivatives  provides the equations determining the eigenvalue $E$, 
which have the same forms for D  and S, 
 \begin{align}
{\rm even:}& \quad \psi^{(+)}(0,E)=\psi^{(-)}(0,E),  \qquad \  \
\frac{d \psi^{(+)}(0,E)}{dx}=0,
 \label{evencond1}\\
 {\rm odd:}& \quad \frac{d \psi^{(+)}(0,E)}{dx}=\frac{d\psi^{(-)}(0,E)}{dx}, \qquad 
 \psi^{(+)}(0,E)=0.  
  \label{oddcond1}
 \end{align}
Thanks to the mirror symmetry,  the first condition is trivially satisfied for both sectors 
by fixing  the relative scales of the $\psi^{(+)}(x)$ and $\psi^{(-)}(x)$,
The second condition determines the eigenvalues $\{E_n\}$ $n=0,1,\ldots$ as  functions of  the system parameters.
In the present case they are  $d$.
The second conditions can be replaced by the equivalent one $\frac{d\psi^{(-)}(0,E)}{dx}=0$ for the even sector
and $\psi^{(-)}(0,E)=0$ for the odd sector.
Obviously the even parity condition \eqref{evencond1} is the {\em Neumann boundary condition} and the
odd parity one \eqref{oddcond1} is the {\em Dirichlet boundary condition}.
This is the rare occasion that the Neumann b.c. appears in quantum mechanics. 
The Dirichlet b.c. appears wherever an impenetrable barrier stands.

Another simplification is built in due to the forms of the double and single well potentials $V_{\text D}(x)$ \eqref{dw1}
and  $V_{\text S}(x)$ \eqref{sw1}.
On the positive half line $x>0$, $V_{\text D}(x)=(x-d)^2$, $V_{\text S}(x)=(x+d)^2$ and they interchange by $d\leftrightarrow -d$.
The same situation happens on the negative half line, $x<0$, too.
Therefore, when the Neumann b.c. equation \eqref{evencond1} is written down for the $V_{\text D}(x)$
wavefunctions, the equation for the $V_{\text S}(x)$ wavefunctions is simply obtained by changing $d$ into $-d$,
and vice versa.  The situation is the same for the Dirichlet b.c. equation \eqref{oddcond1}.
Let us write down the equations determining the eigenvalues $\{E\}$ 
due to the Neumann b.c. \eqref{evencond1} (for the even sector)  and due to 
the Dirichlet b.c. \eqref{oddcond1} (for the odd sector) of the $V_{\text D}$ and $V_{\text S}$ systems as
\begin{align} 
{\rm Double\ well:}\quad &C_{\text D}^{(e)}(d,E)=0,\quad C_{\text D}^{(o)}(d,E)=0,
\label{Dconds}\\
{\rm Single\ well:}\quad &C_{\text S}^{(e)}(d,E)=0,\quad C_{\text S}^{(o)}(d,E)=0,
\label{Sconds}
\end{align}
The following theorem states their close relationship.
 \begin{theo}
 \label{theo:2-1}
 For the even and odd sectors, the functions for the double well and single well are simply related by
 \begin{align} 
{\rm even:}\ C_{\text D}^{(e)}(x,E)=C_{\text S}^{(e)}(-x,E),\qquad
{\rm odd:}\   C_{\text D}^{(o)}(x,E)=C_{\text S}^{(o)}(-x,E), \quad x\in\mathbb{R},
\label{DSrel}
 \end{align}
 up to some  irrelevant constant factors.
 \end{theo}
This is why we call $V_{\text S}(x)$ is the dual potential of $V_{\text D}(x)$, and vice versa.
Without determining these functions, we can safely make the following statement concerning the lower bounds of the eigenvalues of the $V_{\text S}$ system.
\begin{rema}
\label{rem:bound}
For both even and odd sectors, the eigenvalues of the single well system are greater than $d^2$,
\begin{equation}
V_{\text S}(x)\ge d^2 \ \Longrightarrow E>d^2.
\label{Slbound}
\end{equation}
\end{rema}

\section{Polynomial type solutions}
\label{sec:poly}
The most basic result of one-dimensional quantum mechanics is that the Hermite polynomials 
$\{H_n(x)\}$ provide the
complete set of eigenfunctions of the quadratic potential $x^2$. This means,
\begin{align}
 &-\frac{d^2\Psi_{p}(x,2n+1)}{dx^2}+(x+d)^2\Psi_{p}(x,2n+1)=(2n+1)\Psi_{p}(x,2n+1),\n
&\hspace{5cm}\Psi_{p}(x,2n+1)=e^{-(x+d)^2/2}H_n(x+d),\qquad \ n\in\mathbb{Z}_{\ge0},
\label{pher}\\
 &-\frac{d^2\Psi_{m}(x,2n+1)}{dx^2}+(x-d)^2\Psi_{m}(x,2n+1)=(2n+1)\Psi_{m}(x,2n+1),\n
&\hspace{5cm}\Psi_{m}(x,2n+1)=\alpha_ne^{-(x-d)^2/2}H_n(x-d),\quad n\in\mathbb{Z}_{\ge0},
\label{mher}
\end{align}
in which $\alpha_n$ is a constant.
Here the subscript $p$ means `plus' $d$, {\em i.e.} $(x+d)^2$ potential 
and $m$ means `minus' $d$, $(x-d)^2$ potential. 
The degree $n$ Hermite polynomial $H_n(x)$ has  the  parity $H_n(-x)=(-1)^nH_n(x)$.
This means that for $E=2n+1$, for example, $\psi_{\text S}^{(+)}(x,2n+1)=e^{-(x+d)^2/2}H_n(x+d)$
is a piecewise square integrable wavefunction of the single well system on the right half line $x>0$. 
Thus we arrive at a theorem.
\begin{theo}
\label{theo:3-1}
The Neumann \eqref{evencond1} and Dirichlet \eqref{oddcond1} b.c. provide the equations
\begin{align} 
{\rm even:}\ H_n'(d)-dH_n(d)=0,\qquad  {\rm odd:}\ H_n(d)=0,\qquad \bigl(H_n'(d)=2nH_{n-1}(d)\bigr),
\label{Hsconds}
\end{align}
determining a finite number of $\{d\}$'s with which the continuous connection with the left half line 
wavefunction $\psi_{\text S}^{(-)}(x,2n+1)$ is realised.
\end{theo}
Due to the parity of the Hermite polynomial, the contents of these equations are the same 
when $d$ is changed to $-d$, meaning that the above equations apply to the double well system, too.
According to {\bf Theorem \ref{theo:2-1}} we arrive at the following theorem.
\begin{theo}
\label{theo:3-2}
To each positive odd integer $2n+1$ {\rm(}$n\in\mathbb{N}${\rm) }correspond two sets of 
distinct positive parameters
$\{d_j^e\}$, $j=1,\ldots,[\langle\!\langle n\rangle\!\rangle/2]$, and $\{d_j^o\}$, 
$j=1,\ldots,[\langle n\rangle/2]$, satisfying  $H_{n}'(d_j^e)-d_j^eH_n(d_j^e)=0$ and
 $H_n(d_j^o)=0$ \eqref{Hsconds}, respectively.
For the even type $d_j^e$, the Schr\"odinger equations with $V_{\text D}(x)$ and $V_{\text S}(x)$ potential
have an even parity eigenstate with the eigenvalue $2n+1$,
\begin{align} 
\psi_{\text{D},j}^{(e)}(x,2n+1)&=\left\{
\begin{array}{rc}
  e^{-(x+d_j^e)^2/2}H_n(x+d_j^e)&    -\infty<x\le0  \\[2pt]
(-1)^n  e^{-(x-d_j^e)^2/2}H_n(x-d_j^e) &   0\leq x<\infty    
\end{array}
\right.\ , \label{hDevensol}\\[4pt]
\psi_{\text{S},j}^{(e)}(x,2n+1)&=\left\{
\begin{array}{rc}
(-1)^n   e^{-(x-d_j^e)^2/2}H_n(x-d_j^e)&    -\infty<x\le0  \\[2pt]
 e^{-(x+d_j^e)^2/2}H_n(x+d_j^e) &   0\leq x<\infty    
\end{array}
\right..
\label{hSevensol}
\end{align}
For the odd type $d_j^o$, the Schr\"odinger equations with $V_{\text D}(x)$ and $V_{\text S}(x)$ potential
have an odd parity eigenstate with the eigenvalue $2n+1$,
\begin{align} 
\psi_{\text{D},j}^{(o)}(x,2n+1)&=\left\{
\begin{array}{rc}
 - e^{-(x+d_j^o)^2/2}H_n(x+d_j^o)&    -\infty<x\le0  \\[2pt]
(-1)^n  e^{-(x-d_j^o)^2/2}H_n(x-d_j^o) &   0\leq x<\infty    
\end{array}
\right.\ , \label{hDoddsol}\\[4pt]
\psi_{\text{S},j}^{(o)}(x,2n+1)&=\left\{
\begin{array}{rc}
(-1)^n   e^{-(x-d_j^o)^2/2}H_n(x-d_j^o)&    -\infty<x\le0  \\[2pt]
 -e^{-(x+d_j^o)^2/2}H_n(x+d_j^o) &   0\leq x<\infty    
\end{array}
\right..
\label{hSoddsol}
\end{align}
\end{theo}
Here $[a]$ denotes the greatest integer not exceeding $a$ and 
$\langle\!\langle n\rangle\!\rangle=n+1$ for odd $n$ and $\langle\!\langle n\rangle\!\rangle=n$ for even $n$.
Likewise $\langle n\rangle=n+1$ for even $n$ and $\langle n\rangle=n$ for odd $n$.
Obviously these numbers are distinct $\{d_j^e\}\cap\{d_j^o\}=\phi$.
Here we list the explicit expressions of the connection conditions \eqref{Hsconds} 
and the corresponding values of $\{d_j^e\}$ and $\{d_j^o\}$ for $n$  upto 6.
\begin{table}[htp]
\caption{Even parameters}
\begin{center}
\begin{tabular}{|c|c|c|}
\hline
$n$ & $-\bigl(H_n'(d)-dH_n(d)\bigr)$ & $d_j^e$: six digits\\
\hline
1 & $2(-1+d^2)$& 1\\
2& $2d(-5+2d^2)$ & 1.58114\\
3 & $4(3-9d^2+2d^4)$  & 0.602114, 2.03407\\
4 &$4 d (27 - 28 d^2 + 4 d^4)$ & 1.07461, 2.41769\\
5 &$8 (-15 + 75 d^2 - 40 d^4 + 4 d^6)$ &0.476251, 1.47524, 2.75624\\
6& $8 d (-195 + 330 d^2 - 108 d^4 + 8 d^6)$ & 0.881604, 1.82861, 3.06251\\
\hline
\end{tabular}
\end{center}
\label{tab:evenpara}
\end{table}%
\begin{table}[htp]
\caption{Odd parameters}
\begin{center}
\begin{tabular}{|c|c|c|}
\hline
$n$ & $H_n(d)$ & $d_j^o$: six digits \\
\hline
2& $4 (-1 + 2 d^2)$ & 0.707107\\
3 & $24 d (-3 + 2 d^2)$  & 1.22474\\
4 &$96 (3 - 12 d^2 + 4 d^4)$ & 0.524648, 1.65068\\
5 &$960 d (15 - 20 d^2 + 4 d^4)$ &0.958572, 2.02018\\
6& $5760 (-15 + 90 d^2 - 60 d^4 + 8 d^6)$ &0.436077, 1.33585, 2.3506\\
\hline
\end{tabular}
\end{center}
\label{tab:oddpara}
\end{table}%
It is straightforward to verify that the even connection condition \eqref{Hsconds} is the same as (I.21) of
Znojil's paper \cite{znojil22} and the odd condition \eqref{Hsconds} agrees with (I.24).

The upper bound of the zeros of the Hermite polynomial $H_n(x)$ is known \cite{szego} (6.32.6),
\begin{equation}
H_n(x_j^{(n)})=0,\qquad x_j^{(n)}<\sqrt{2n+1}-\frac{c}{(2n+1)^{1/6}},\quad c=1.85575 \ldots .
\label{hebound}
\end{equation}
This means that the eigenvalues of these explicitly known {\em odd} states are greater than 
the corresponding $(d_j^o)^2$,
\begin{equation}
(d_j^o)^2<2n+1,
\label{eigbound}
\end{equation}
which is consistent with {\bf Remark \ref{rem:bound}}.
We do not know  a corresponding bound for the even sector, that is the zeros of $H_n'(d)-dH_n(d)=0$.

We would not call these exactly solvable states QES (quasi-exactly solvable states) \cite{ushv}.
A quantum mechanical system with a quasi-exactly solvable potential has a finitely 
many exactly solvable states. In most cases these states are related by $sl(2,R)$  algebra \cite{turb}.
In the present case, a double \eqref{dw1} or single well \eqref{sw1} potential with the parameter $d$
being the zeros of \eqref{Hsconds} has only one exactly solvable state. This is a totally different situation from QES.

We will come back to the topic of the integer eigenvalues in the second half of the subsequent section.
Before closing this section, let us emphasise the fact that   the above connection conditions \eqref{Hsconds}, 
the Neumann and Dirichlet b.c. including the fact that they are identical for the double and single well systems,
are intuitively quite easy to understand.
\section{Non-polynomial exact eigenfunctions}
\label{sec:nonpoly}
\subsection{Confluent hypergeometric functions}
\label{sec:kummer}

It  is well known that 
the one dimensional Schr\"odinger equation with the quadratic potential $x^2$ 
can be rewritten as an equation of the confluent hypergeometric function $\varphi(z,E)$,
\begin{align}
 &-\frac{d^2\psi(x,E)}{dx^2}+x^2\psi(x,E)=E\psi(x,E),\quad \psi(x,E)=e^{-x^2/2}\varphi(z,E),\quad z\eqdef x^2,\n
&\hspace{3cm}\Longrightarrow z\frac{d^2\varphi(z,E)}{dz^2}+(b-z)\frac{d\varphi(z,E)}{dz}-a\varphi(z,E)=0,
\label{kummereq}\\
& \hspace{3cm} a=\frac{(1-E)}4,\quad b=\frac12.
\label{abdef}
\end{align}
The two fundamental solutions of the above Kummer's differential equation \eqref{kummereq} are
\begin{align} 
\varphi_1(z,E)&={}_1F_1(a,b\,;z)=\sum_{k=0}^\infty\frac{(a)_k}{(b)_k}\frac{z^k}{k!},
\label{fsol1}\\
\varphi_2(z,E)&=z^{1-b}{}_1F_1(a+1-b,2-b\,;z)=z^{1-b}\sum_{k=0}^\infty\frac{(a+1-b)_k}{(2-b)_k}\frac{z^k}{k!},
\end{align}
in which $(a)_n$ is the shifted factorial,
\begin{equation}
(a)_n\eqdef\frac{\Gamma(a+n)}{\Gamma(a)}=\prod_{k=0}^{n-1}(a+k)=a(a+1)\cdots(a+n-1).
\end{equation}
The  well-known piecewise  square integrable combination of the fundamental solutions is
\begin{equation}
U(a,b\,;z)\eqdef \frac{\Gamma(1-b)}{\Gamma(a-b+1)}{}_1F_1(a,b\,;z)
+\frac{\Gamma(b-1)}{\Gamma(a)}z^{1-b}{}_1F_1(a-b+1,2-b\,;z).
\label{Udef}
\end{equation}
Since $b=\tfrac{1}{2}$, $\Gamma(-\tfrac{1}{2})=-2\Gamma(\tfrac{1}{2})=-2\sqrt{\pi}$, we introduce
\begin{equation}
\bar{U}(a,\tfrac{1}{2}\,;z)\eqdef \frac{1}{\Gamma(a+\tfrac{1}{2})}{}_1F_1(a,\tfrac{1}{2}\,;z)
-\frac{2}{\Gamma(a)}z^{1/2}{}_1F_1(a+\tfrac{1}{2},\tfrac{3}{2}\,;z),
\label{Ubdef}
\end{equation}
for simplicity of presentation.
For the present case, corresponding to the two types of $\sqrt{z}$ as a function of $x$, 
$\sqrt{z_p}=\pm(x+d)$, $\sqrt{z_m}=\pm(x-d)$, we choose  the branch
of $\Psi_{p}(x,E)$ and $\Psi_{m}(x,E)$ in such a way  $\sqrt{z_p}>0$ and $\sqrt{z_m}>0$  at 
infinity, so that the wavefunctions are damped at plus and minus infinity,
\begin{align}
 -\frac{d^2\Psi_{p}(x,E)}{dx^2}&+(x+d)^2\Psi_{p}(x,E)=E\Psi_{p}(x,E),\quad z_p\eqdef(x+d)^2,\n
 \Psi_{p}(x,E)&=
\left\{
\begin{array}{cc}
\Psi_{p}^{(+)}(x,E)=e^{-z_p/2}\bar{U}_{p}^{(+)}(a,\tfrac{1}{2}\,;z_p) &   x\ge 0,\  \sqrt{z_p}=x+d, \\[3pt]
\Psi_{p}^{(-)}(x,E)=\alpha e^{-z_p/2}\bar{U}_{p}^{(-)}(a,\tfrac{1}{2}\,;z_p)  &     x\le0,  \ \   \sqrt{z_p}=-(x+d),
\end{array}
\right., 
\label{vpphianaly}\\[4pt]%
 -\frac{d^2\Psi_{m}(x,E)}{dx^2}&+(x-d)^2\Psi_{m}(x,E)=E\Psi_{m}(x,E),\quad z_m\eqdef(x-d)^2,\n
 \Psi_{m}(x,E)&=
\left\{
\begin{array}{cc}
\Psi_{m}^{(+)}(x,E)=e^{-z_m/2}\bar{U}_{m}^{(+)}(a,\tfrac{1}{2}\,;z_m) &   x\ge 0,\ \sqrt{z_m}=x-d,   \\[3pt]
\Psi_{m}^{(-)}(x,E)=\beta e^{-z_m/2}\bar{U}_{m}^{(-)}(a,\tfrac{1}{2}\,;z_m)  &   x\le 0, \   \sqrt{z_m}=d-x,
\end{array}
\right..
\label{vmphianaly}
\end{align}
in which $\alpha$ and $\beta$ are constants to be determined later.
 Within the interval $-d<x<d$, which include the connection point $x=0$, the wavefunctioms
 $\psi_{\text D}^{(\pm)}(x,E)$ and   $\psi_{\text S}^{(\pm)}(x,E)$ are expressed 
 by
 \begin{align} 
 \psi_{\text D}^{(+)}(x,E)&=\Psi_{m}^{(+)}(x,E),\quad  \psi_{\text D}^{(-)}(x,E)=\Psi_{p}^{(-)}(x,E),
 \label{Dwav}\\
  \psi_{\text S}^{(+)}(x,E)&=\Psi_{p}^{(+)}(x,E),\quad  \psi_{\text S}^{(-)}(x,E)=\Psi_{m}^{(-)}(x,E).
 \label{Swav}
 \end{align}
The boundary values of the double well wavefunctions are
\begin{align*} 
e^{d^2/2}\Psi_{\text D}^{(+)}(0,E)&
=\frac{1}{\Gamma(a+\tfrac{1}{2})}{}_1F_1(a,\tfrac{1}{2}\,;d^2)
+\frac{2d}{\Gamma(a)}{}_1F_1(a+\tfrac{1}{2},\tfrac{3}{2}\,;d^2),\\
e^{d^2/2}\Psi_{\text D}^{(+)'}(0,E)&= \frac{d}{\Gamma(a+\tfrac{1}{2})}
\left\{{}_1F_1(a,\tfrac{1}{2}\,;d^2)-2{}_1\dot{F}_1(a,\tfrac{1}{2}\,;d^2)\right\}\\
&  -\frac{2}{\Gamma(a)}\left\{(1-d^2)\cdot{}_1F_1(a+\tfrac{1}{2},\tfrac{3}{2}\,;d^2)
+2d^2{}_1\dot{F}_1(a+\tfrac{1}{2},\tfrac{3}{2}\,;d^2)\right\},
\end{align*}
in which
\begin{equation}
{}_1\dot{F}_1(a,b\,;z)\eqdef\frac{d\,{}_1F_1(a,b\,;z)}{dz}=\frac{a}{b}\cdot{}_1F_1(a+1,b+1\,;z).
\label{Fdifdef}
\end{equation}
According to {\bf Theoren \ref{theo:2-1}} the corresponding quantities for the single well wavefunctions 
are obtained by changing $d$ to $-d$, 
These lead to the following theorem.
\begin{theo}
\label{theo:4-1}
The connection conditions for the double and single well wavefunctions are
\begin{align} 
{\rm D, even:}& \quad \alpha=1,\quad \frac{d}{\Gamma(a+\tfrac{1}{2})}
\left\{{}_1F_1(a,\tfrac{1}{2}\,;d^2)-2\,{}_1\dot{F}_1(a,\tfrac{1}{2}\,;d^2)\right\}\n
&\qquad  \qquad -\frac{2}{\Gamma(a)}\left\{(1-d^2)\cdot{}_1F_1(a+\tfrac{1}{2},\tfrac{3}{2}\,;d^2)
+2d^2{}_1\dot{F}_1(a+\tfrac{1}{2},\tfrac{3}{2}\,;d^2)\right\}=0,
\label{UDevencond}\\
{\rm D, odd:}& \quad \alpha=-1,\quad \frac{1}{\Gamma(a+\tfrac{1}{2})}{}_1F_1(a,\tfrac{1}{2}\,;d^2)
+\frac{2d}{\Gamma(a)}{}_1F_1(a+\tfrac{1}{2},\tfrac{3}{2}\,;d^2)=0.
\label{UDoddcond}\\
{\rm S, even:}& \quad \beta=1,\quad \frac{d}{\Gamma(a+\tfrac{1}{2})}
\left\{{}_1F_1(a,\tfrac{1}{2}\,;d^2)-2\,{}_1\dot{F}_1(a,\tfrac{1}{2}\,;d^2)\right\}\n
&\qquad  \qquad +\frac{2}{\Gamma(a)}\left\{(1-d^2)\cdot{}_1F_1(a+\tfrac{1}{2},\tfrac{3}{2}\,;d^2)
+2d^2{}_1\dot{F}_1(a+\tfrac{1}{2},\tfrac{3}{2}\,;d^2)\right\}=0,
\label{USevencond}\\
{\rm S, odd:}& \quad \beta=-1,\quad \frac{1}{\Gamma(a+\tfrac{1}{2})}{}_1F_1(a,\tfrac{1}{2}\,;d^2)
-\frac{2d}{\Gamma(a)}{}_1F_1(a+\tfrac{1}{2},\tfrac{3}{2}\,;d^2)=0.
\label{USoddcond}
\end{align}
The zeros of these equations provide the eigenvalues  $\{E_n\}$ 
of the $V_{\rm D}$ and $V_{\rm S}$ systems in the even and odd sectors. 
The corresponding eigenfunctions are those listed in \eqref{Dwav} and \eqref{Swav}.
\end{theo}
\begin{rema}
\label{rem:otherbr}
Roughly speaking, the functions in the $V_{\rm S}$ system \eqref{USevencond},\eqref{USoddcond}
are the other branches of the corresponding $\dot{U}$ and $U$ functions in the $V_{\rm D}$
system \eqref{UDevencond},\eqref{UDoddcond} and vice versa.
\end{rema}

We show the seven lowest eigenvalues, 4 from the even sector and 3 from the odd sector,
of the double ({Table \ref{tab:Dweig}}) and single ({Table \ref{tab:Sweig}}) well potentials
for a selected small values of $d$. For each value of $d$, the eigenvalues of the odd sectors are
greater than the corresponding ones in the even sectors, as dictated by the oscillation theorem.
Each specific eigenvalue of the $V_{\text S}$ system increases monotonically with the parameter $d$.
The $d$-dependence of the states of the $V_{\text D}$ system is quite interesting.
As $d$ increases above 3, the split between the even and odd sectors diminishes appreciably.
One could say for almost safely that when $d$ increases the tunneling effects of the lowest $n$
($n< d^2/2$) states in each sector disappear.  This would mean that  for large $d$,
the eigenvalues of $V_{\text D}(x)$
approach to $E_n^e=1+2n-\epsilon_n$, $E_n^o=1+2n+\epsilon_n'$ with very small $\epsilon_n,\epsilon_n'>0$.
It is a good challenge to find out the asymptotic behaviours of $E_n^e(d)$ and $E_n^o(d)$ of the 
$V_{\text S}(x)$ system.
\begin{table}[h]
\caption{7 lowest eigenvalues of $V_{\text D}$}
\begin{center}
\begin{tabular}{|c|c|c|c|c|c|c|c|}
\hline
d& $E_0^e$ & $E_0^o$ &$E_1^e$ &$E_1^o$ &$E_2^e$ & $E_2^0$ & $E_3^e$ \\
\hline
0 & 1 & 1 & 3& 3 & 5 & 5& 7 \\
1/10 & 0.895426& 2.78209&4.72612 &6.66950 & 8.62731& 10.5849& 12.5497\\
1/4 & 0.768973& 2.48392 & 4.34603& 6.20358&8.09868 &9.99237  & 11.9046\\
1/2&0.635529 &2.06077 & 3.79417& 5.50548&7.29817 &9.08421  & 10.9098\\
3/4 &0.590301 & 1.72471&3.34471 &4.90343 &6.59770 &8.27404 & 10.0146\\
1& 0.618919&1.46847 &3 & 4.39493&5.99720 & 7.56038&9.21846 \\
3/2&0.801494 &1.15748 &2.64868 &3.64627 &5.10400& 6.41679& 7.92382\\
2 &0.951419 & 1.03576& 2.73504& 3.22301&4.67082 &5.64089 & 7.04349\\
3 &0.999551 &1.00039 & 2.99252&3.00604 &4.94552&5.03982 &6.79866 \\
4 & 0.999999&1.000000 & 2.99998& 3.00001& 4.99977&5.00020 &6.99802 \\
\hline
\end{tabular}
\end{center}
\label{tab:Dweig}
\end{table}%

\begin{table}[h]
\caption{7 lowest  eigenvalues of $V_{\text S}$}
\begin{center}
\begin{tabular}{|c|c|c|c|c|c|c|c|}
\hline
d& $E_0^e$ & $E_0^o$ &$E_1^e$ &$E_1^o$ &$E_2^e$ & $E_2^0$ & $E_3^e$ \\
\hline
1/10 &1.12121 &3.23353 &5.29034 &7.34657& 9.38899& 11.4312& 13.4665\\
1/4&1.33487 &3.61368 & 5.75688&7.89681&10.0032 &12.1086 & 14.1970\\
1/2 & 1.77790& 4.32871&6.61797 &8.89589 &11.1096 & 13.3194& 15.4967\\
3/4 &2.33218 & 5.14812& 7.58472& 9.99898&12.3203 &14.6339 & 16.9002\\
1 & 3&6.07439 & 8.65856& 11.2076&13.6366 &16.0533 &18.4086 \\
3/2 & 4.68276&8.25537 &11.1329 &13.9472 &16.5907 &19.2113 & 21.7441\\
2 & 6.83597& 10.8843&14.0506 &17.1244 &19.9803 & 22.8017&25.5108 \\
5/2 &9.46595 & 13.9704& 17.4196&20.7471 & 23.8127& 26.8318&29.7154 \\
\hline
\end{tabular}
\end{center}
\label{tab:Sweig}
\end{table}%

A few remarks on the numerical calculations of the eigenvalues. 
For $z\to +\infty$, $U(a,b\,;z)$ 
\eqref{Udef} behaves asymptotically $\sim z^{-a}$. 
The expressions in  {\bf Theorem \ref{theo:4-1}} \eqref{UDevencond}--\eqref{USoddcond} 
increase drastically $\sim d^{E/2}$ as $d$ and $E$ increase. 
With certain reduction factors  the zeros (the eigenvalues) of these expressions can be determined 
as precisely as wanted for a specified  parameter $d$. 
This preciseness propagates to the preciseness of the eigenfunctions. 
This is why  the systems with $V_{\text D}(x)$ and $V_{\text S}(x)$  
belong to the category of potentials of non-polynomial exact solvability \cite{znojil22}.

\begin{figure}[h]
    \begin{tabular}{cccc}
      \begin{minipage}[b]{0.45\hsize}
        \centering
        \includegraphics[keepaspectratio, scale=0.8]{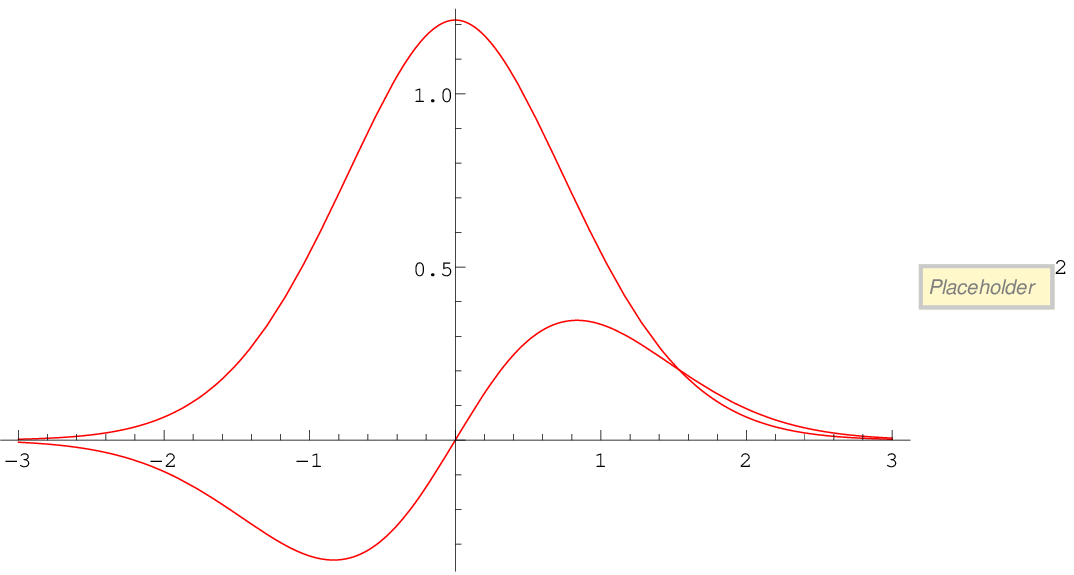}
        \caption{$d=1$, $E=3,\ 6.07$, S}
        \label{fig:1}
      \end{minipage} &
      \begin{minipage}[b]{0.45\hsize}
        \centering
        \includegraphics[keepaspectratio, scale=0.8]{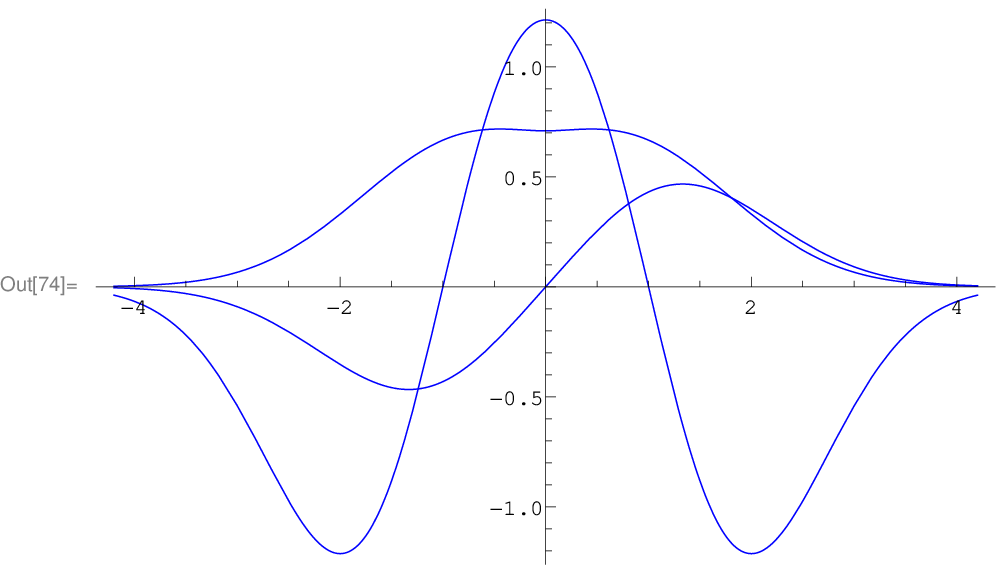}
        \caption{$d=1$, $E=0.619, 1.468,\ 3$, D}
        \label{fig:2}
      \end{minipage}
         \end{tabular}
 \end{figure}

\begin{figure}[h]
    \begin{tabular}{cccc}
      \begin{minipage}[c]{0.45\hsize}
        \centering
        \includegraphics[keepaspectratio, scale=0.8]{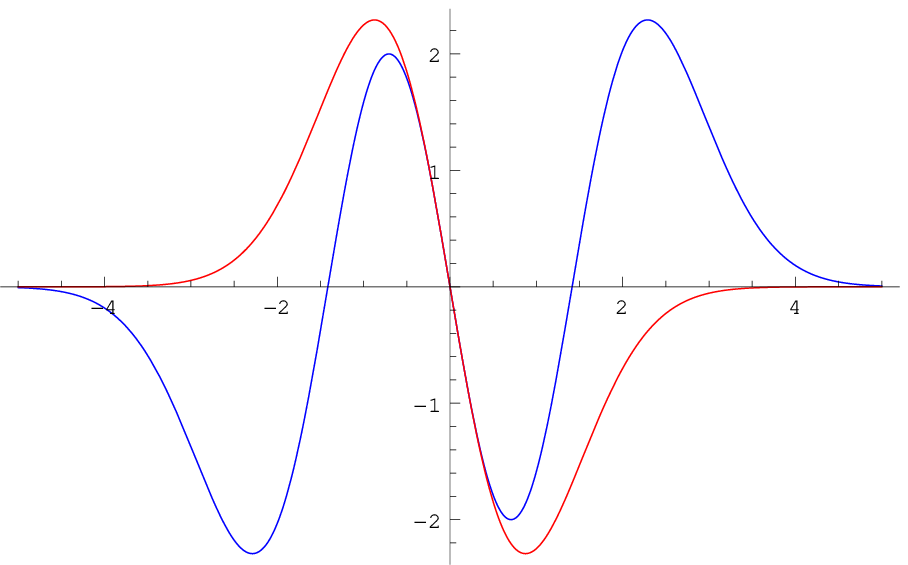}
        \caption{$E=5,d=1/\sqrt{2}$ odd}
        \label{fig:3}
      \end{minipage} &
      \begin{minipage}[c]{0.45\hsize}
        \centering
        \includegraphics[keepaspectratio, scale=0.8]{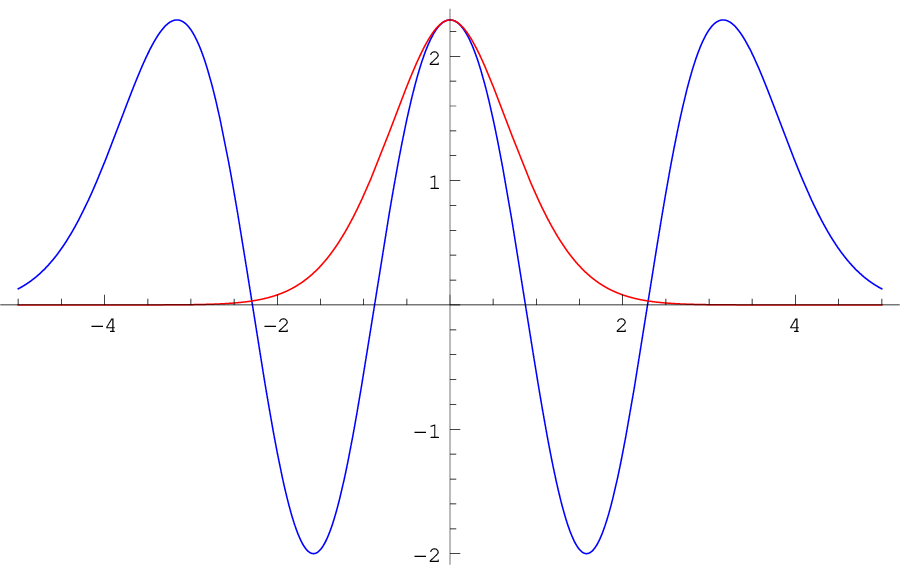}
        \caption{$E=5, d=\sqrt{5/2}$ even}
        \label{fig:4}
      \end{minipage} 
    \end{tabular}
  \end{figure}

In order to share the vidid images of polynomial and non-polynomial type  eigenfunctions of the low lying eigenstates
we present four figures. The two lowest eigenfunctions  of the $V_{\text S}(x)$ 
potential with $d=1$ are shown in Fig.\ref{fig:1},
 the ground state $E_0^e=3$ and the first excited state $E_0^o=6.07439$.
For comparison, we show the three lowest energy  states of the $V_{\text D}$ 
system with $d=1$, $E_0^e=0.618919$  $E_0^o=1.46846$ and $E_1^e=3$  in Fig.\ref{fig:2}. 
The two lowest ones  are not of  the polynomial type. 
Two  $E=5$ eigenstates in the odd sector with $d=1/\sqrt{2}$ are shown in Fig.\ref{fig:3}
and those in the even sector with $d=\sqrt{5/2}$ are displayed in Fig.\ref{fig:4}.
Those of the $V_{\text S}$ system have red lines and those in $V_{\text D}$ blue.


Fig.\ref{fig:2} shows the effect of splitting due to tunneling.
Let us introduce a pair of potentials restricted to the right and left half line by an infinite impenetrable wall
at the origin,
\begin{equation}
V_{\rm DR}(x)=
\left\{
\begin{array}{cc}
(x-d)^2  & x>0     \\
+\infty  &   x=0  
\end{array}
\right.,
\qquad 
V_{\rm DL}(x)=
\left\{
\begin{array}{cc}
+\infty  &   x=0  \\
(x+d)^2  & x<0     \\
\end{array}
\right.,\qquad d>0.
\label{VDRLpot}
\end{equation}
They have the same set of eigenvalues. The   eigenfunctions are  restricted to the right and left line
satisfying the Dirichlet b.c. at the origin.
The function vanishing on the left half  line and take the odd line of Fig.\ref{fig:2} at $x>0$ is the
ground state eigenfunction of $V_{\rm DR}(x)$ with $E=1.46846>1$, $d=1$, 
since $V_{\rm DR}(x)|_{d=1}>(x-1)^2$ on the left half line.
Likewise the vanishing on the right half line and take the odd line of Fig.\ref{fig:2} at $x<0$ is the
ground state eigenfunction of $V_{\rm DL}(x)$ with $E=1.46846$, $d=1$.
When the infinite barrier is removed and $V_{\rm DR}(x)$ and $V_{\rm  DL}(x)$ 
merge to become $V_{\text D}(x)$, 
the odd combination of these states becomes the first excited states with
the same eigenvalue $E=1.46846$. The removal of the  infinite barrier at the origin could be 
rephrased as the addition of an infinitely deep and narrow well at the origin. 
This has no effect on the odd combination of the original ground state eigenfunctions as they vanish at the origin.
However, this has the effect of increasing the even combination of the 
wavefunctions at the origin to the point of satisfying the 
Neumann b.c.  and thus decreasing the eigenvalue. 
This mechanism applies to all the eigenfunctions of the $V_{\rm DR}(x)$ and $V_{\rm DL}(x)$
systems.
Therefore, the splitting of the eigenlevels in the $V_{\text D}(x)$ potential 
means pushing down  the even states whereas the odd states stay 
at the original eigenvalues of $V_{\rm DL}$ and $V_{\rm DR}$.
The situation is essentially the same for the $V_{\text S}(x)$ potential. Instead of \eqref{VDRLpot}
\begin{equation}
V_{\rm SR}(x)=
\left\{
\begin{array}{cc}
(x+d)^2  & x>0     \\
+\infty  &   x=0  
\end{array}
\right.,
\qquad 
V_{\rm SL}(x)=
\left\{
\begin{array}{cc}
+\infty  &   x=0  \\
(x-d)^2  & x<0     \\
\end{array}
\right.,\qquad d>0.
\label{VSRLpot}
\end{equation}
are the potentials to be considered.

\subsection{Revisiting polynomial type solutions}
\label{sec:poly2}
Since the parameter $a$ in the Kummer's differential equation \eqref{kummereq} is $a=(1-E)/4$ \eqref{abdef},
various quantities and expressions in the previous subsection 
simplify a lot for the odd integer eigenvalues  $E$,
\begin{equation}
E=4n+1,\ 4n+3, \ \Longleftrightarrow a=-n,\ a+\tfrac12=-n,\, \qquad n\in\mathbb{Z}_{\ge0}.
\label{intE}
\end{equation}
For these values the confluent hypergeometric functions ${}_1F_1(a,b\,;z)$ and 
${}_1F_1(a+\tfrac12,b\,;z)$  terminate and become polynomials in $z$.
Many expressions in \S\ref{sec:kummer} reduce to those in \S\ref{sec:poly}. Here we list them for comparison.

\paragraph{(i) $E=4n+1$} This means 
\begin{equation*}
a=-n\  \Rightarrow \frac{1}{\Gamma(a)}=\frac1{\Gamma(-n)}=0, 
\quad  U(-n,\tfrac12\,;z)=\frac{\Gamma(\tfrac12)}{\Gamma(-n+\tfrac12)}{}_1F_1(-n,\tfrac12\,;z).
\end{equation*}
Here ${}_1F_1(-n,\tfrac12\,;x^2)$ is a degree $n$ polynomial in $x^2$,
\begin{equation}
{}_1F_1(-n,\tfrac12\,;x^2)=\frac{n!}{(\tfrac12)_n}L_n^{(-\tfrac12)}(x^2),\qquad 
H_{2n}(x)=(-1)^nn!2^{2n}L_n^{(-\tfrac12)}(x^2),
\end{equation}
in which $L_n^{(-\tfrac12)}(x)$ is the Laguerre polynomial.
\paragraph{(ii) $E=4n+3$} This means 
\begin{equation*}
a+\tfrac12=-n\  \Rightarrow \frac{1}{\Gamma(a+\tfrac12)}=\frac1{\Gamma(-n)}=0, 
\quad  U(-n-\tfrac12,\tfrac12\,;z)=\frac{\Gamma(-\tfrac12)}{\Gamma(-n-\tfrac12)}\sqrt{z}\,{}_1F_1(-n,\tfrac32\,;z).
\end{equation*}
Here ${}_1F_1(-n,\tfrac32\,;x^2)$ is a degree $n$ polynomial in $x^2$,
\begin{equation}
{}_1F_1(-n,\tfrac32\,;x^2)=\frac{n!}{(\tfrac32)_n}L_n^{(\tfrac12)}(x^2),\qquad 
H_{2n+1}(x)=(-1)^nn!2^{2n+1}xL_n^{(\tfrac12)}(x^2),
\end{equation}
By using these relations, one can easily verify that \eqref{UDevencond}--\eqref{USoddcond} reduce to 
\eqref{Hsconds}.
\section{Summary and Comments}
\label{sec:comm}
Some basic facts, expressions and numbers  related with the eigenvalues and eigenfunctions 
of the piecewise analytic and exactly solvable
potentials $V_{\text D}(x)$ and $V_{\text S}(x)$ are explored.
For the applications, the norms of some of the lower lying eigenstates would be needed. 
This would require a substantial work.

It is well known that a piecewise linear potential $V_{\text L}(x)=g^3|x|$ is exactly solvable. 
It is expected that the potentials 
\begin{equation}
V_{\rm LD}(x)=\text{min}[V_{\text L}(x+d),V_{\text L}(x-d)],\qquad 
V_{\rm LS}(x)=\text{max}[V_{\text L}(x+d),V_{\text L}(x-d)],
\label{linpot}
\end{equation}
would be exactly solvable 
by the same procedures as those used in this paper.
More interesting would be the double and single well versions of a Krein-Adler deformation \cite{krein,adler, os29}
of the harmonic oscillator potential,
\begin{align}
&\hspace{4cm}V_{\rm KA}(x)=x^2+3+\frac{32x^2}{(2x^2+1)^2}-\frac{8}{2x^2+1},
\label{dubovpot}\\
&V_{\rm KAD}(x)=\text{min}[V_{\rm KA}(x+d),V_{\rm KA}(x-d)],\quad
V_{\rm KAS}(x)=\text{max}[V_{\rm KA}(x+d),V_{\rm KA}(x-d)],\quad d>0.
\label{DSdubov}
\end{align}
Dubov et al \cite{dubov} introduced the  exactly solvable $V_{\rm KA}(x)$, which has the regular singular points at 
$x=\pm \frac{i}{\sqrt{2}}$ with the characteristic exponents $(-1,2)$. 
The complete set of the eigenvalues and eigenfunctions are
\begin{equation}
{\psi}_{\rm{KA},n}(x)=\frac{e^{-x^2/2}\text{W}[H_1,H_2,H_n](x)}{4(2x^2+1)}, 
\quad \mathcal{E}(n)=2n,\quad n\in\mathbb{Z}_{\ge0}\backslash\{1,2\},
\label{adlerH2fun}
\end{equation}
in which 
\begin{align}
&\text{W}\,[f_1,\ldots,f_m](x)
  \eqdef\det\Bigl(\frac{d^{j-1}f_k(x)}{dx^{j-1}}\Bigr)_{1\leq j,k\leq m},
  \label{wron}
\end{align}
is the Wronskian of functions $\{f_1,\ldots,f_m\}$. At least we can find the polynomial 
type eigenfunctions of $V_{\rm KAD}(x)$ and $V_{\rm KAS}(x)$ quite easily.  
If these $V_{\rm KAD}(x)$ and  $V_{\rm KAS}(x)$  turn out to be exactly solvable,
we would have an infinitely many similar potentials by the Krein-Adler prescriptions \cite{os29}.

\section*{Acknowledgements}
R.\,S. thanks Milosh Znojil for sending \cite{znojil22} just after publication.

\bigskip
\noindent
{\bf Informed Consent Statement}: Not applicable. \\
{\bf Data Availability Statement}: No new  data created\\
{\bf Conflicts of Interest}: There is no conflict of interests. 


\begin{thebibliography}{99}
\bibitem{znojil22}
M.\, Znojil, ``Displaced harmonic oscillator  $V\sim \text{min}[(x+d)^2,(x-d)^2]$ as a benchmark double well
potential," Quantum Rep. {\bf 4} (2022) 309-323.


\bibitem{sz1}
R.\,Sasaki and M.\,Znojil,
``One-dimensional Schr\"{o}dinger equation with non-analytic potential
$V(x)= -g^2\exp (-|x|)$ and its exact Bessel-function solvability,"
J. Phys. {\bf A49} (2016)  445303 (12pp),
{\tt arXiv:1605.07310[math-ph]}. 

\bibitem{sz2}
M.\,Znojil, ``Symmetrized exponential oscillator,"  Mod. Phys. Lett. {\bf A31} (2016) 1650195;\\
R.\,Sasaki,  ``Confining non-analytic exponential potential $V(x) = g^2{\rm exp}(2|x|)$ 
and its exact Bessel-function solvability. 
{\tt arXiv:1611.02467}.

\bibitem{sz3}
M.\, Znojil, ``Morse potential, symmetric Morse potential and bracketed bound-state energies,"
Mod. Phys. Lett. {\bf A31} (2016) 1650088;\\
R.\,Sasaki, ``Symmetric Morse potential is exactly solvable," {\tt arXiv:1611.05952}.


\bibitem{infhul}
L.\,Infeld and T.\,E.\,Hull,
``The factorization method,''
Rev. Mod. Phys. {\bf 23} (1951) 21-68.

\bibitem{susyqm}
F.\,Cooper, A.\,Khare and U.\,Sukhatme,
``Supersymmetry and quantum mechanics,''
Phys. Rep. {\bf 251} (1995) 267-385.

\bibitem{genden}
L.\,E.\,Gendenshtein,
``Derivation of exact spectra of the Schroedinger equation by means of
supersymmetry,''
JETP Lett. {\bf 38} (1983) 356-359.

\bibitem{gomez}
D.\,Gomez-Ullate, N.\,Kamran and R.\,Milson,
``An extended class of orthogonal polynomials defined by a Sturm-Liouville
problem,''
J. Math. Anal. Appl. {\bf 359} (2009) 352-367, {\tt arXiv:0807.3939[math-ph]};
``An extension of Bochner's problem: exceptional invariant sub-spaces,''
J. Approx. Theory {\bf 162} (2010) 987-1006, {\tt arXiv:0805.3376[math-ph]}.

\bibitem{quesne}
C.\,Quesne,
``Exceptional orthogonal polynomials, exactly solvable potentials and
supersymmetry,''
J. Phys. A: Math. Theor. {\bf 41} (2008) 392001 (6 pp),
{\tt arXiv:0807.4087\hspace{0pt}[quant-ph]}.

\bibitem{os16}
S.\,Odake and R.\,Sasaki,
``Infinitely many shape invariant potentials and new orthogonal polynomials,''
Phys. Lett. {\bf B679} (2009) 414-417,
{\tt arXiv:0906.0142[math-ph]}.

\bibitem{os25}
S.\,Odake and R.\,Sasaki,
``Exactly solvable quantum mechanics and infinite families of multi-indexed
orthogonal polynomials,''
Phys. Lett. {\bf B702} (2011) 164-170,
{\tt arXiv:1105.\hspace{0pt}0508[math-ph]}.

\bibitem{os31}
S.\, Odake and R.\, Sasaki,
``Non-polynomial extensions of solvable potentials
\'a la Abraham-Moses,"
J. Math. Phys. {\bf 54} (2013) 102106 (19pp),
{\tt arXiv:1307.0910\hspace{0pt}[math-ph]}.

\bibitem{krein}
M.\,G.\,Krein,
``On continuous analogue of a formula of Christoffel from the theory
of orthogonal polynomials," (Russian)
Doklady Acad. Nauk. CCCP, {\bf 113} (1957) 970-973.

\bibitem{adler}
V.\,\'E.\,Adler,
``A modification of Crum's method,''
Theor. Math. Phys. {\bf 101} (1994) 1381-1386.


\bibitem{os29}
S.\, Odake and R.\, Sasaki,
``Krein-Adler transformations for shape-invariant potentials and pseudo virtual states,"
J. Phys. A {\bf 46} (2013) 245201 (24pp)
{\tt arXiv:1212.6595\hspace{0pt}[math-ph]}

\bibitem{dubov}
S.\,Yu.\,Dubov, V.\,M.\,Eleonski\u{i} and N.\,E.\,Kulagin,
``Equidistant spectra of anharmonic oscillators,''
 Soviet Phys. JETP {\bf 75} (1992) 446-451;
 ``Equidistant spectra of anharmonic oscillators,''
Chaos {\bf 4} (1994) 47-53.

\bibitem{szego}
G.\,Szeg\"o, {\it Orthogonal Polynomials}, Amer. Math. Soc.
Colloquium Publications Vol. 23 (Amer. Math. Soc., New York,
1939).

\bibitem{ushv}
A.\,G.\, Ushveridze,  {\it  Quasi-Exactly Solvable Models in Quantum Mechanics},  IOPP: Bristol, UK, (1994).

\bibitem{turb}
A.\,V.\,Turbiner, ``Quasi-exactly solvable problems and sl(2) algebra,"  Commun. Math. Phys.  {\bf 118} (1988) 467-474.

\goodbreak
\end{thebibliography}
\end{document}